\documentclass[runningheads,a4paper]{llncs}

\usepackage{amssymb}
\usepackage{graphicx}

\newcommand{\keywords}[1]{\par\addvspace\baselineskip
\noindent\keywordname\enspace\ignorespaces#1}

\begin{document}

\mainmatter

\title{A real world network pricing game with less severe Braess' Paradox }

\titlerunning{Network pricing game with less severe Braess' Paradox}

\author{Abhimanu Kumar
\thanks{IIT Guwahati, India}%
\and Sanjib Kumar Das%
\thanks{IIT Kharagpur, India}%
}
\authorrunning{Kumar and Das}

\institute{}

\toctitle{Reducing Braess' Paradox For A Real World Network Pricing game}
\tocauthor{Kumar and Das}
\maketitle

\begin{abstract}

Internet and graphs are very much related. The graphical structure of internet has been studied extensively to provide efficient solutions to routing and other problems. But most of these studies assume a central authority which controls and manages the internet. In the recent years game theoretic models have been proposed which do not require a central authority and the users are assumed to be routing their flows selfishly. The existence of Nash Equilibria, congestion and the amount of inefficiency caused by this selfish routing  is a major concern in this field. A type of paradox in the selfish routing networks, Braess' Paradox, first discovered by Braess, is a major contributor to inefficiency. Several pricing mechanisms have also been provided which give a game theoretical model between users(consumers) and ISPs ({Internet Service Providers} or sellers) for the internet.\newline

We propose a novel pricing mechanism, based on real world Internet network architecture, which reduces the severity of Braess' Paradox in selfish routing game theoretic networks. It's a pricing mechanism between combinatorial users and ISPs. We prove that Nash equilibria exists in this network  and provide bounds on inefficiency . We use graphical properties of internet to prove our result. Several interesting extensions and future work have also been discussed.

\end{abstract}

\section{Introduction}

Internet has grown meteorically in the few decades and thus the amount of study associated with it.  The graphical and combinatorial properties of internet have been studied in great deal. Several interesting graphical properties have been derived with experimental results \cite{ref1} . The earlier studies assumed a central authority for the internet for ease of analysis . But today's Internet is more of an autonomous system without any central authority where users and the ISPs  work selfishly to maximize their own interests. Because of this behavior there has been growing concerns about QoS (Quality of Service), finding efficient routes, pricing mechanisms etc. Since each user or ISP is selfish and works for his own interest without any concern for the overall efficiency of the system, a game theoretic approach to this problem looks to suit much better. Based on these lines there has been a growing amount of research literature in theoretical computer science about analysis of the inefficiency due to selfishness of the agents (users and ISPs) of the system. Several models have been suggested \cite{ref2,ref3}. Roughgarden et. al \cite{ref4} study the inefficiency arising due to Braess' Paradox, first discovered by Braess \cite{ref5} and later reported by Murchland \cite{ref6}. To combat the ills arising due to selfishness like congestion etc. several pricing mechanisms have  been suggested in\cite{ref7,ref8,ref9} but the effectiveness of these rely on the owner of the resources i.e the ISP. The ISPs are selfish and their goals may not align with social objectives of efficiency and QoS. There is also vast amount of research which proposes pricing mechanisms so that the resources of the ISPs can be effectively sold to the users\cite{ref10,ref11} keeping in mind the selfish behavior. But all of these either assume a constant cost function for the edges charged by ISPs or analyze their models for cases when each user routes a negligible amount of flow for ease of analysis, or assume a kind of coordination between different users.

In selfish routing games the main aim of the user is to minimize its latency cost (delay). It does not worry about the cost charged by the ISPs. Several researches \cite{ref4,ref8} have studied this kind of game giving bounds on Price of Anarchy which measure the inefficiency arising due to selfishness. In Network Design games the main aim for the users is to have a network which will let them flow their required flows with some pricing mechanism, it's not concerned about congestion delays. This game has also been the focus of attention of several works \cite{ref2,ref3}. Here the job of the ISPs is to just provide the network with the minimal price, generally a constant cost per edge, without themselves(ISPs) having any selfish motives. It's just the user who has selfish motives. Obviously none of the above mentioned games capture the full complexity of the network. We provide a model which takes both the ISPs and the users to be selfish and the user cost function has both the factors: the latency as well as cost charged by the ISPs, which is not always constant.

We give a model which captures a real world general case network scenario, where the price of the edges,charged by ISPs, in the network varies with amount of flow. The idea behind this assumption is that the per unit flow cost charged by seller decreases if the buyer buys more. i.e. if you buy in bulk you pay less per unit cost . Our cost function also includes the latency factor that is latency caused due to congestion in the network edge. In this model we assume combinatorial users (buyers) and their flow is not negligible. Each user has a significant amount of flow . Also we assume that the flow is non-splittable. This complicates the situation a lot. Our model can easily be extended for negligible and splittable flow cases as well. We prove that the effect of Braess' paradox is less severe in our model. We give a better bound of the worst case of Braess' Paradox. We also prove that the  Nash Equilibria exists in our model.

\subsection{Related Work}

Hayrapetyan et. al \cite{ref10} analyze among similar lines as us but for their cost function they assume a constant term for the per unit flow price charged by ISPs but in the real world scenario generally the per unit charge decreases if the flow required by the user increases i.e. ISPs provide concession for more flow. Our model captures this. They have not done any analysis of the effect of  Braess' Paradox in their network, too. We show that in our model the severity of Braess' Paradox is reduced.

\subsection{Model and Notations}

Our network is a \emph{multicommodity flow network} i.e. it has more than one \emph{source-target} pair. This model is similar to the model discussed in \cite{ref12} with an added term for unit cost function for the price charged by ISPs. Let us consider a directed graph \emph{G = (V,E)} where \emph{V} is the set of vertices and \emph{E} is the set of Edges. There is a set of \emph{k} source and target(sink) pair of vertices $(s_1,t_1),(s_2,t_2)...( s_{k},t_{k} )$  and a set of \emph{k} users who want to use this network for their flow between a $(s_{i},t_{i})$ pair. Each $(s_i,t_i)$ has a corresponding user \emph{i} and vice-versa . Each user \emph{i} has a flow $f_i > 0$ .Different players can have identical source sink pair. $\Pi_i$  is used to denote the set of $(s_{i},t_{i})$ paths of the network for a given \emph{i}. Each user for $(s_{i},t_{i})$ pair picks a path P from the set $\Pi_i$ for its flow and thus $f_{P}^{i} > 0$ and equals 0 for all other paths. Thus the strategy set for user \emph{i} is $\Pi_i$. The condition that the flow is 0 for all other paths is termed as the non-splittable flow routing. And as we see the number of users is combinatorial , this condition is termed as Atomic Selfish Routing \cite{ref12}. A flow \emph{f} is a feasible flow for this network if it corresponds to a strategy profile : for each player \emph{i} , $f_P^i$ equals $r_i$ for exactly one path P for $s_i-t_i$  where $r_i$ is the total amount of flow player \emph{i} has to send.

\begin{figure}
\centering
\includegraphics[width=8.2cm]{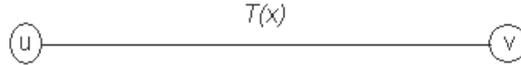}
\caption{shows the cost function of the edge between the vertices \emph{u} and \emph{v}}
\label{fig:genGraph}
\end{figure}

\textbf{$\Pi$} is the union of all the sets $\Pi_i$ i.e. \textbf{$\Pi$} = $\bigcup_{i=1}^k \Pi_i$. Every edge \emph{e} has cost \emph{$t_e^{i}(f_e)$} for each user \emph{i}, $t_e^{i} : R^+ \rightarrow R^+$, and \emph{$f_e$} is the total flow amount in the edge \emph{e} i.e. \textbf{$f_e = \sum\limits_{e \epsilon P : P \epsilon  \Pi : i=1} ^ k f_{P}^{i} $}. The cost function \emph{$t_e^{i}$} contains contains two terms. One is a term for congestion C(f) and another is a term U(f) that ISPs charge for the flow of user i $f_{P}^{i}$ in that edge where $e \epsilon P$. Thus $t_e(f_e)_{i} = C(f_e) + U(f_P^{i}) = c_e(f_e) + u_e(f_P^{i})$ for user \emph{i} for using edge \emph{e}. Note that $c_e()$ is the $e_{th}$ component of Function C() or $c_e()$ is the \emph{e}'s contribution to the cost function C().We can say that $T_e(x)$ is the total cost of all the user for edge \emph{e} i.e. $t_e(x) = \sum\limits_{i=1}^k t(x_e)_i$.Note that $t_e(x_e)_{i}$ will be 0 for users \emph{i} for edge \emph{e} who dont route their flow through \emph{e}. The first component of the cost function $t_e{i}$ is defined as $c_e : R^+ \rightarrow R^+$ . This is dependent upon the total flow \emph{x} in the edge . We consider only those $C_e$ which are convex with respect to flow \emph{x} . So for the time being lets take $C_e(x) = x $. The Second term of the cost function $t_e{i}$ is defined as $u_e^{i} : R^+ \rightarrow R^+$ . This function is the per unit flow charge of the ISP for the edge. This function will be of the form $F(f_i)/f_i$ where $F(f_i)$ is the charge for the total flow $f_i$ of the user \emph{i}  by the ISP and thus the per unit cost of routing the flow for the user \emph{i} will be $F(f_i)/f_i$. You can note that the congestion term is independent of user \emph{i} , because even if the user doesn't contribute to the congestion it confronts the congestion of the edge caused by all the users.

Now we define the concept of equilibria in this model. As we know the users are selfish so they will all try to minimize their cost . Let us define the cost of a path \emph{P} with respect to a flow \emph{f} in terms of the sum of the costs for constituent edges: $t_P^{i}(f) = \sum\limits_{e \epsilon P} t_e^{i}(f_e) = \sum\limits_{e \epsilon P} (c_e(f_e) + u_e(f_P^{i} ))$.

\begin{definition}\label{equilibriumCond} (\textbf{Equilibrium Flow}){  Let f be a feasible flow for the network G(V,E) . The flow f is an equilibrium flow if, for every player $i\epsilon\{1,2,....,k\}$ and every pair  $P, \tilde{P} \epsilon \Pi_i $ of $s_i-t_i$ paths with $f_P^i >0$
\begin{equation}
t_P^i(f) \leq t_{\tilde{P}}^i(\tilde{f})
\end{equation}
where $\tilde{f}$ is the flow identical to f except that $\tilde{f}_P^i = 0$ and $\tilde{f}_{\tilde{P}}^i = r_i$ }
\end{definition}

Now we define the Social cost in our network. Social cost is the total cost experienced by all the users in the network. i.e. $\sum\limits_{i=1}^k (\texttt{total cost of each user i})$ Formally Social Cost \emph{\textbf{SC(f) }}for the flow \emph{f}  is defined as

\begin{definition} (\textbf{Social Cost}){  Given the network instance (G,r,t) where $t_e = c_e+ u_e$ (i.e. cost has two components) and $r_i$ is the amount of flow with each user i with f the flow in the network. Then social cost \emph{\textbf{SC(f)}} is
\begin{equation}\label{costSocial}
SC(f) = \sum\limits_{e\epsilon E} c_e(f_e)f_e + \sum\limits_{i=1}^k \sum\limits_{e\epsilon P_i} u_e(r_i)r_i
\end{equation}
 }
\end{definition}

Informally when you calculate the sum of the cost of all the k users in the network we arrive at equation (\ref{costSocial}) .Next we  define \emph{Price Of Anarchy}  of the network. This term was defined for \emph{Social Cost} of the network and is used to measure inefficiency arising in the network flow due to selfish users . It was first conceptualized by Papadimitriou et.al \cite{ref2} and Anshelevich et.al \cite{ref3} Our definition is the same as used by Roughgarden et.al in  \cite{ref12}.

\begin{definition} (\textbf{Price of Anarchy}){  In the network instance (G,r,t) as above Price of Anarchy \textbf{POA} is the ratio of the social cost in the worst case equilibrium flow f and that of the optimal flow $f^{*}$ .So  \begin{equation}\textbf{POA}=\frac{SC(f)}{SC(f^{*})} \end{equation}
}
\end{definition}\label{POA}


\subsection{Our Results}

\begin{enumerate}

    \item We show that for our model there always exists a pure Nash Equilibria even in the case of Combinatorial Users with non-splittable flows in real world network scenarios. \newline

    \item We show that the severity of Braess' Paradox is reduced in our case . The previous bound on Braess' Paradox was 4/3 given by Roughgarden et. al. \cite{ref4} .Our bound is 8/7 in the worst case. We also show a trivial result that if the Unit Cost function term is the dominant term compared to the Congestion Term in the Cost Function then the Braess' Paradox doesn't even exist for the original graph given by Braess \cite{ref5} which is also used by Roughgarden for the 4/3 bound.\newline

    \item We give bound for Price of Anarchy for our model

\end{enumerate}

\subsection{Paper Organization}

From here on we discuss the existence of pure equilibria in our model. Then in the next section we prove the reduced severity of Braess' Paradox. After that We give bounds on Price of Anarchy  for our model. In the last section we conclude our work with some open problems and opportunity for future work.

\section{Pure Equilibria Existence In the Model}

We prove the existence of pure equilibria in our model using the  potential functions. This method was suggested in \cite{ref12}. Our potential function is different from theirs .Potential function is a function which captures the changes in the cost of network very effectively when any user deviates from a network state or flow condition. We will use a discrete combinatorial potential function. The idea behind this combinatorial function is that it has finite values so there must exist a minimum value among those values. This minimum is what we are interested in. We make our potential function such that it depicts the change experienced by a user deviating from the equilibria. And hence we prove that we have attained a state of equilibria.In the following theorem we are going to define the proof of the existence of equilibria.

\begin{theorem} (Equilibrium in the network) {Let (G,r,t) is a network instance where every user i has an amount $r_i$ to flow and the cost function is t as defined in earlier sections in the given graph G. Then (G,r,t) has at least one equilibrium flow given that the congestion factor in the t is an affine function.
}
\end{theorem}

\textbf{proof} Note that we took that the congestion factor in the \emph{t}  an affine function as an affine function is a fair estimate of congestion in a real world network. We design a potential function to capture our network model. As defined in the previous section our cost function \emph{t} has  two components. Let us formally define our cost function. The cost function $t_e^i$ is the amount of per unit flow cost experienced by user \emph{i} by flowing $r_i$ amount of flow  on edge \emph{e} which lies on path \emph{P} chosen by i out of $\Pi_i$. $t_e^i = c_e(f_e) + u_e(r_i)$ . Here as defined earlier $f_e = \sum\limits_{i \epsilon S_e} r_i$ . Where $S_e$ is the set of users whose path has edge e. Now we define our Potential function \textbf{$\Phi$} as
\begin{equation}
\Phi = \sum\limits_{e \epsilon E} \left( c_e(f_e)f_e+ \sum\limits_{i \epsilon S_e} c_e(r_i)r_i \right) + \sum\limits_{i \epsilon S_e}u_e(r_i)r_i
\end{equation}

Having defined the potential function for our network we use its features to prove our result. Since the network instance has finite users and each user has finite strategies, thus there are only finite number of flows and correspondingly finite number of values for our potential function. Say for a flow f the potential function has the minimum value. We prove that this corresponds to the equilibrium flow. We prove it by contradiction. Let us say that this flow f is not the equilibrium flow. That means by switching its path a user i can strictly decrease his cost. Let the previous path was P for that user and the new path is $\tilde{P}$ . Now change in the cost of i is:

\begin{equation}\label{usrCostChange}
0 > t_{\tilde{P}}(\tilde{f}) - t_P(f) = \sum\limits_{e \epsilon \tilde{P}\backslash P } \left(c_e(f_e+r_i) + u_e(r_i) \right) - \sum\limits_{e \epsilon P\backslash\tilde{P} } \left(c_e(f_e)+ u_e(r_i) \right)
\end{equation}

Now let us look at the change in the potential function because of the change in user \emph{i}'s strategy. As we know that $c_e(f_e)$ is an affine function so it can be written as $a_{e}f_e+b_e$
$$
  \Phi = \sum\limits_{e \epsilon E} \left(\underline{ c_e(f_e)f_e+ \sum\limits_{i \epsilon S_e} c_e(r_i)r_i }\right) + \sum\limits_{i \epsilon S_e}u_e(r_i)r_i
$$

Rewriting the underlined term in the above equation we get
\begin{eqnarray*}
 c_e(f_e)f_e+ \sum\limits_{i \epsilon S_e} c_e(r_i)r_i & = & (a_{e}f_e + b_e)f_e+ \sum\limits_{i \epsilon S_e} (a_{e}r_i + b_e)r_i \\
 &=& \left(a_e\sum\limits_{i \epsilon S_e} r_i + b_e\right)\sum\limits_{i \epsilon S_e} r_i +\sum\limits_{i \epsilon S_e} (a_{e}r_i^2 + b_{e}r_i) \\
 &=& a_e(\sum\limits_{i \epsilon S_e} r_i)^2 + 2b_e\sum\limits_{i \epsilon S_e} r_i +a_e\sum\limits_{i \epsilon S_e} r_i^2 \\
 &=& 2a_e\sum\limits_{i \epsilon S_e} r_i^2 + 2a_e\sum\limits_{i,j \epsilon S_e , i\neq j}r_{i}r_{j} + 2b_e\sum\limits_{i \epsilon S_e} r_i
\end{eqnarray*}

So $\Phi$ can also be written

\begin{eqnarray*}
\Phi = \sum\limits_{e \epsilon E} \left( 2a_e\sum\limits_{i \epsilon S_e} r_i^2 + 2a_e\sum\limits_{i,j \epsilon S_e , i\neq j}r_{i}r_{j} + 2b_e\sum\limits_{i \epsilon S_e} r_i \right) + \sum\limits_{i \epsilon S_e}u_e(r_i)r_i
\end{eqnarray*}

Now the change $\Delta\Phi$ in the potential function because of the change in user \emph{i}'s strategy is:

\begin{eqnarray*}
\Delta\Phi &=& \sum\limits_{e \epsilon \tilde{P}\backslash P }\left(2a_{e}r_i^2 + 2a_{e}r_{i}f_{e} + 2b_{e}r_{i} + u_{e}(r_i)r_i\right) - \sum\limits_{e \epsilon \tilde{P}\backslash P }\left(2a_{e}r_i^2 +  2a_{e}r_{i}(f_{e}-r_i) + 2b_{e}r_{i} + u_{e}(r_i)r_i\right)\\
&=& 2r_i\left[\sum\limits_{e \epsilon \tilde{P}\backslash P }\left( a_e(f_e+r_i)+b_e + u_{e}(r_i) \right) - \sum\limits_{e \epsilon \tilde{P}\backslash P }\left( a_e(r_i+f_e-r_i)+b_e + u_{e}(r_i)\right)\right]
\end{eqnarray*}

\begin{equation}\label{potFuncChange}
\Longrightarrow \Delta\Phi=2r_i\left[\sum\limits_{e \epsilon \tilde{P}\backslash P } \left(c_e(f_e+r_i) + u_e(r_i) \right) - \sum\limits_{e \epsilon P\backslash\tilde{P} } \left(c_e(f_e)+ u_e(r_i) \right)\right]
\end{equation}

The bracketed term in the right hand side of this equation is same as equation (\ref{usrCostChange}). Thus this means that $\Delta\Phi < 0$ which contradicts our assumption that f gives the minimum value for $\Phi$. Thus f is the equilibrium flow for instance (G,r,t).\newline \qed

\section{Braess' Paradox with  Reduced Severity}
\begin{figure}
\centering
  \begin{tabular}{cc}
    \includegraphics[width=2.5in]{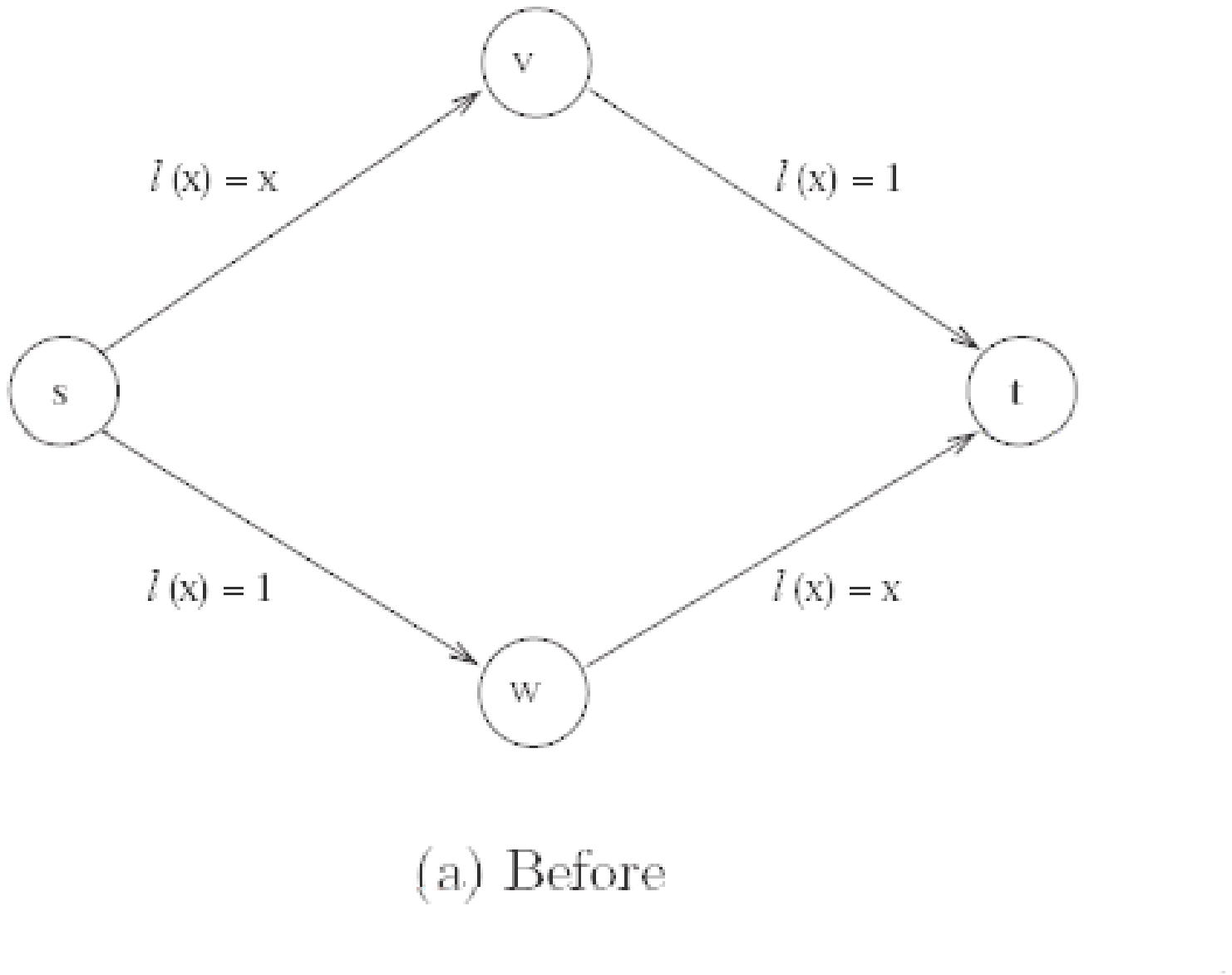}
    &
    \includegraphics[width=2.5in]{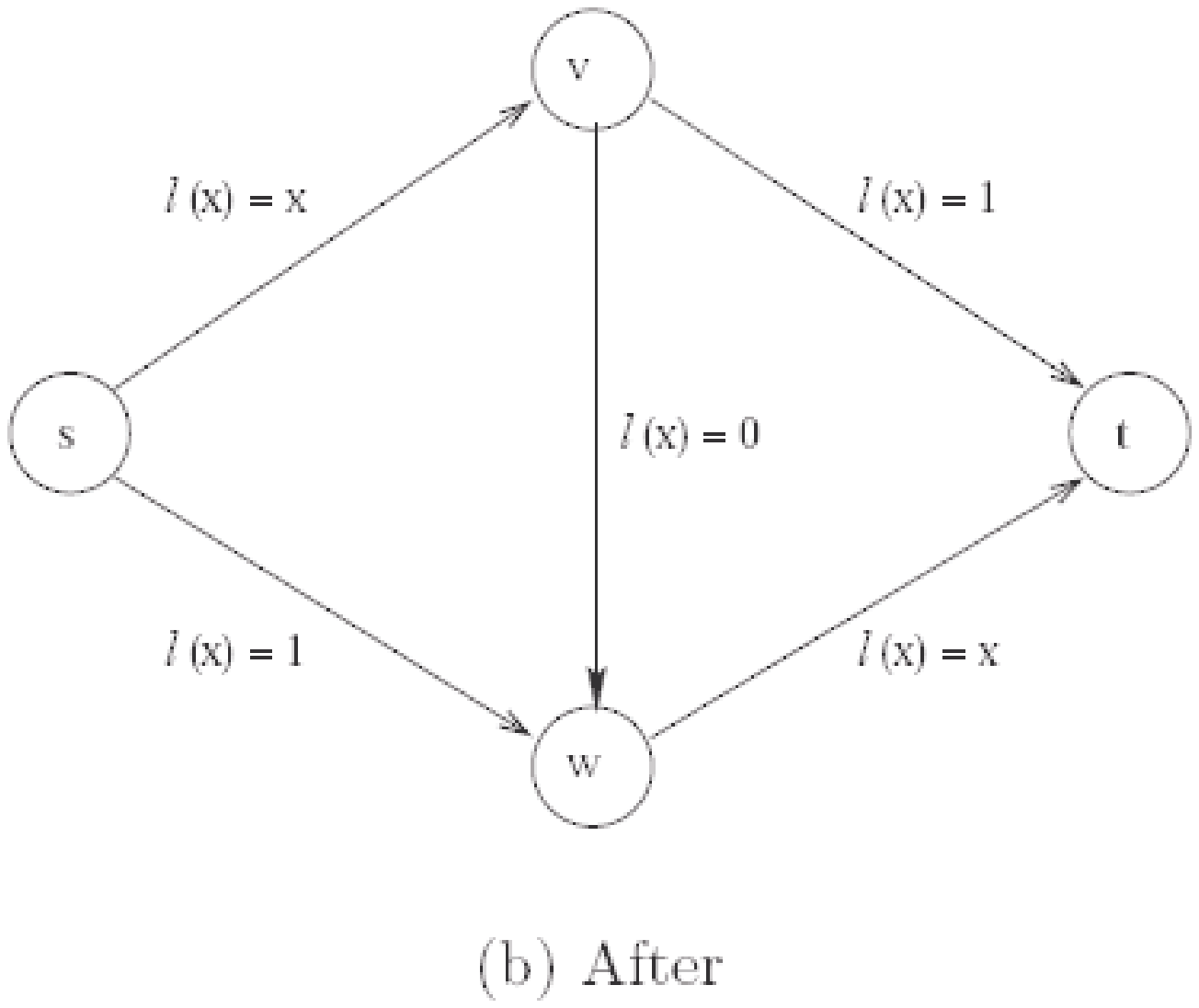}
  \end{tabular}

\caption{Effect of Braess' Paradox in the Original network}
\label{fig:braessOldNw}
\end{figure}

In this section we show that our network model reduces the severity of Braess' Paradox. Braess' Paradox is most conspicuous in the following graph, fig(\ref{fig:braessOldNw}).

 This graph is used by Roughgarden et.al \cite{ref12} to give the 4/3 bound of anarchy. Assuming that the total  flow of all the users  is one unit and there are N users where N is very large,in the graph in  fig(\ref{fig:braessOldNw}a), at equilibrium the flow is evenly distributed between the two paths i.e.each path has 1/2 units of flow. Thus the cost experienced by each user i is 3/2 per unit flow. We feel intuitively that addition of the zero cost directed edge $v\rightarrow w$ should decrease the cost , but when we add the path any user \emph{i} uses the path $u\rightarrow v\rightarrow w\rightarrow t$, because at any stage if the flow in the path $u\rightarrow v\rightarrow w\rightarrow t$ was \emph{x} then user \emph{i} experiences \emph{2x} unit cost which is less than \emph{1+x} unit cost in the other paths, But since every user is selfish each one migrates to this path and thus \emph{x} becomes 1.This makes the per unit cost experienced by each user 2. thus the ratio is $\frac{2}{3/2}$ i.e. 4/3. This is the worst case effect of Braess' Paradox. This was the bound i.e. that any addition of edge can increase the network cost by atmost 4/3.

\begin{figure}
\centering
  \begin{tabular}{cc}
    \includegraphics[width=2.5in]{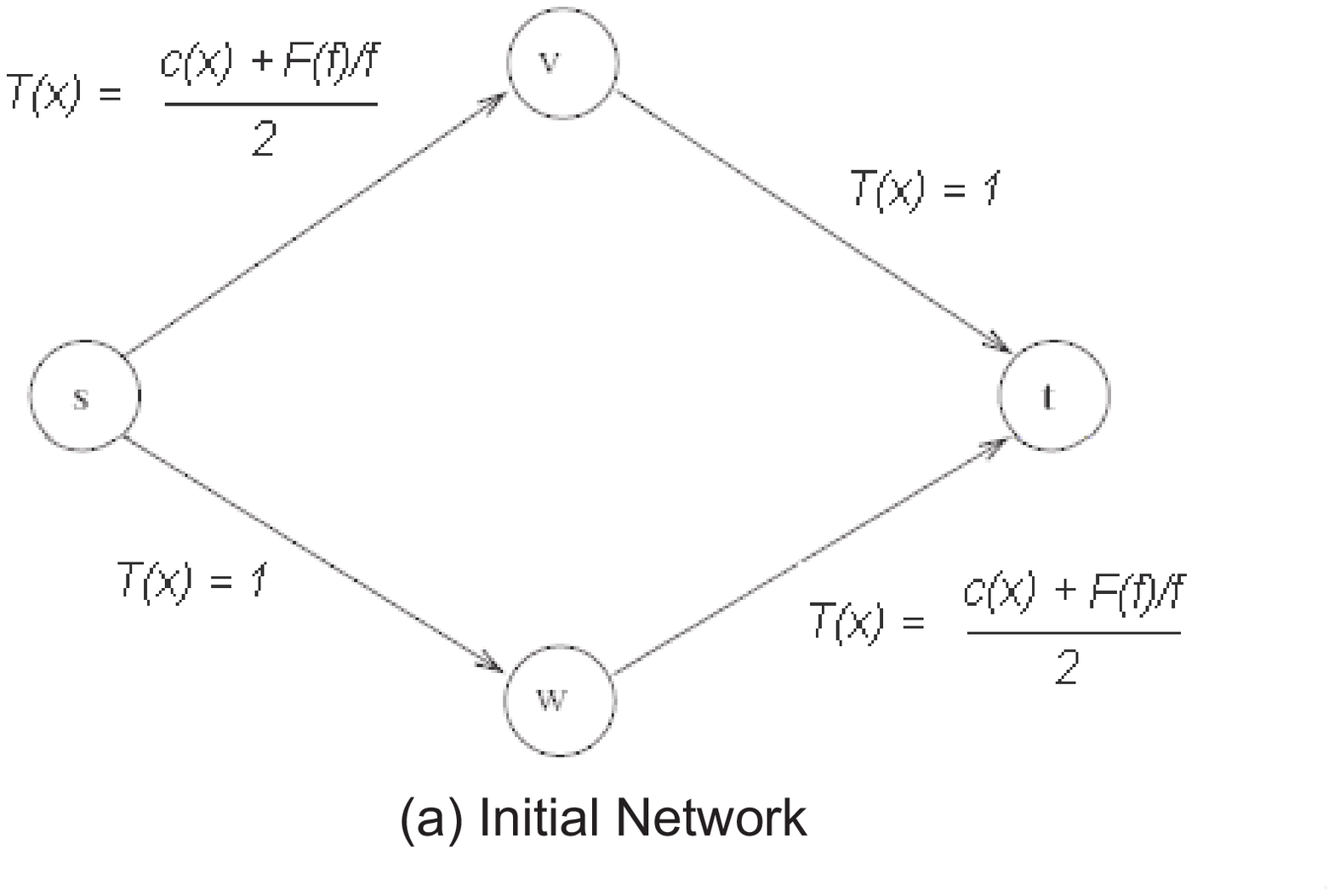}
    &
    \includegraphics[width=2.5in]{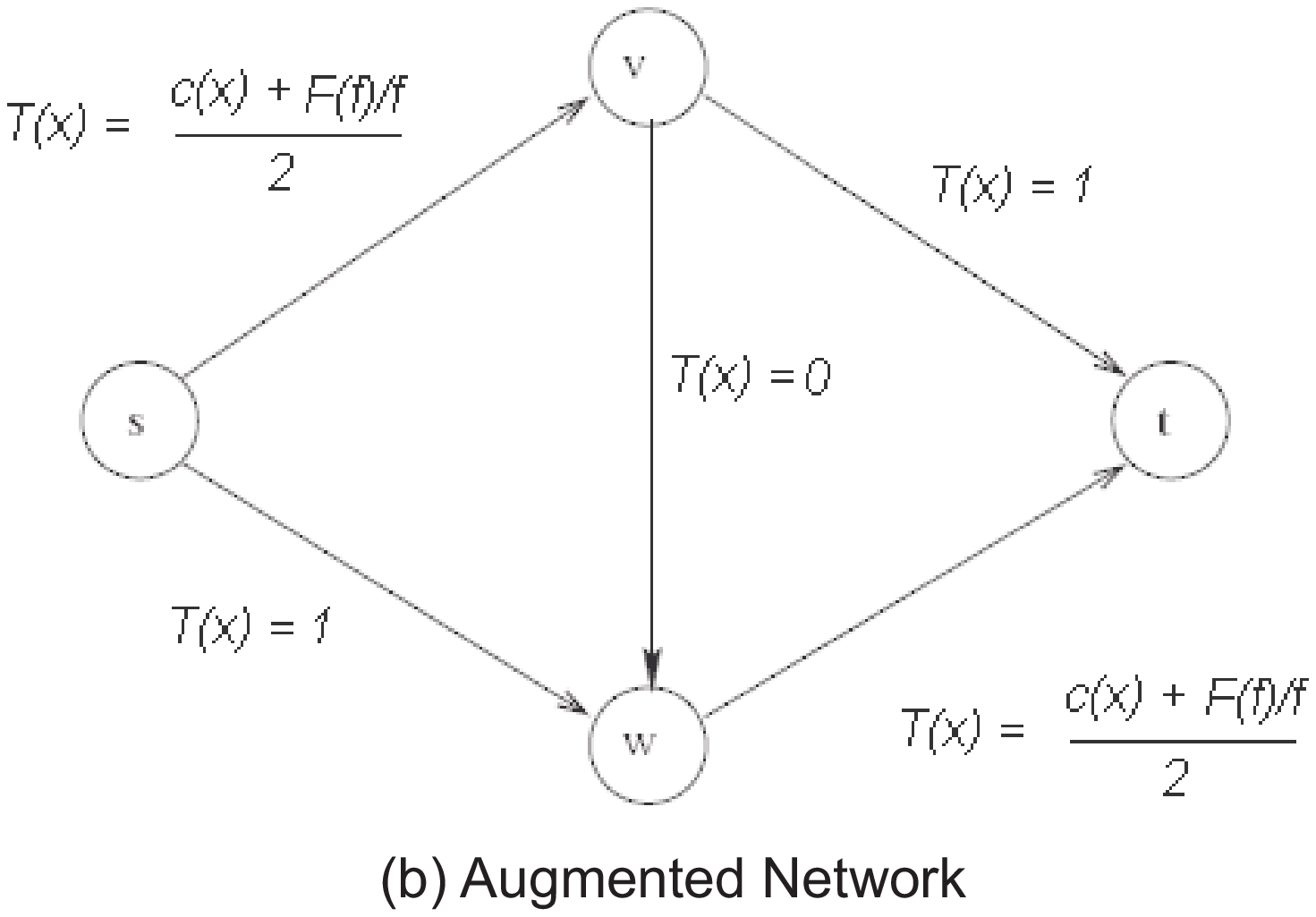}
  \end{tabular}

\caption{Effect of Braess' Paradox in the new network}
\label{fig:braessNewNw}
\end{figure}

Now let us consider our network in fig(\ref{fig:braessNewNw}).

\begin{figure}
\centering
\includegraphics[width=8.2cm]{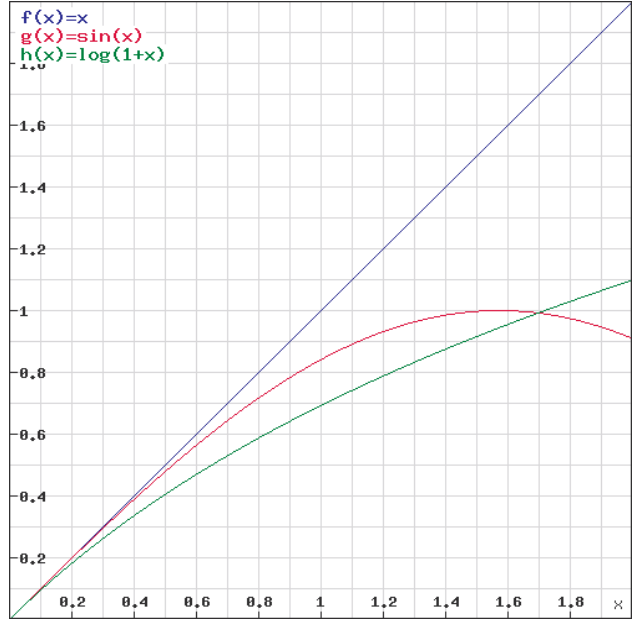}
\caption{Unit Cost function plot}
\label{fig:unitCostGraph}
\end{figure}

In fig(\ref{fig:braessNewNw}a) we have cost function $t_e^i = \frac{c_e(f_e) + u_e(f^i)}{2}$ where $u_e(f^i) = \frac{F_e(f^i)}{f^i}$ and $F_e(f^i)$ is the cost charged by ISP from user i for flowing $f^i$ units. The logic behind using this cost component $u_e(f^i) = \frac{F_e(f^i)}{f^i}$ is that as the user purchase more flow in the network edge the ISP gives it discounts. So the effective $u_e(f^i)$, per unit cost flow, decreases with the increase in flow. This is what happens in real world network pricing. We take the congestion cost term $c_e(x)$ to be a linear function i.e. $c_e(f_e)=f_e$. First lets have a look at the unit cost function $u_e(x) = F(x)/x$. Since the whole network is normalized i.e. total flow is 1, cost for edges $s\rightarrow w$ and $v\rightarrow t$ is 1 there is no loss of generality in assuming that for a flow $\varepsilon$, $F(\varepsilon)$ is $\varepsilon$ where $\varepsilon$ is very small i.e $lim_{ \varepsilon \to 0} F(\varepsilon) = \varepsilon$ and the value of $F(\varepsilon + \delta\varepsilon) < \varepsilon + \delta\varepsilon $ i.e. $\lim_{x \to 0} ({u(x)= \frac{F(x)}{x}}) = 1$ . We have plotted the graph of some such functions like $\sin{x}$ and $\log{(1+x)}$  in fig(\ref{fig:unitCostGraph}) against graph $y=x$ for comparing how the marginal cost for a flow $\delta{X}$ decreases as the X increases. So we have \begin{equation} \label{unitCostInequality}
x\geq 0 \Rightarrow ({u(x)= \frac{F(x)}{x}}) \leq 1
 \end{equation}

Now suppose there were N users each with 1/N units of flow with the total flow in the network 1 unit, then in the graph of fig(\ref{fig:braessNewNw}a) the total flow of 1 unit will be divided into the two paths equally with each user flowing 1/N units and the per unit cost experienced by any user i is $1+ \frac{c_e(1/2)+u_e(1/N)}{2} = 1+(\frac{\frac{1}{2} + \frac{F_e(1/N)}{1/N}}{2})$. Now with the addition of the zero cost directed edge in fig(\ref{fig:braessNewNw}b)  any user i uses the path $u\rightarrow v\rightarrow w\rightarrow t$, because at any stage if the flow in the path $u\rightarrow v\rightarrow w\rightarrow t$ was x then user i experiences 2($\frac{x + \frac{F_e(1/N)}{1/N}}{2}$) unit cost which is less than 1+($\frac{x + \frac{F_e(1/N)}{1/N}}{2}$) unit cost in the other paths as shown below:

From Equation (\ref{unitCostInequality}) we have
\begin{eqnarray*}
1/N \geq 0 \Rightarrow \frac{F_e(1/N)}{1/N} \leq 1 \\
\Rightarrow  \frac{F_e(1/N)}{1/N} + x \leq 1 + 1 = 2 \texttt{ As }x \leq 1 \\
\Rightarrow  2(\frac{F_e(1/N)}{1/N} + x) \leq  2 +\frac{F_e(1/N)}{1/N} + x \\
\Rightarrow 2(\frac{x + \frac{F_e(1/N)}{1/N}}{2}) \leq 1+(\frac{x + \frac{F_e(1/N)}{1/N}}{2})
\end{eqnarray*}

That is what we wanted as the result. So the user i chooses path $u\rightarrow v\rightarrow w\rightarrow t$ but since every user is selfish each one chooses this path and thus increasing cost. Thus at equilibrium the cost to the user is $(1 + \frac {F_e(1/N)}{1/N})$ and the ration between the new cost and the old cost i.e. bound $\rho$ is\begin{equation}\label{theRho}\rho = \frac{(1 + \frac {F_e(1/N)}{1/N})}{(1+(\frac{\frac{1}{2} + \frac{F_e(1/N)}{1/N}}{2}))} = \frac {4 + 4\frac {F_e(1/N)}{1/N})}{5 + 2\frac {F_e(1/N)}{1/N})}\end{equation}
Since $\frac {F_e(1/N)}{1/N}) \leq 1 $ , thus the worst case value of $\rho$ is $\frac{4+4}{5+2} = 8/7$ \newline\qed

Now let us have a normalized generic $t_e$ that is $t_e = c_{1}c_e + c_{2}u_e$ where $c_1+c_2 = 1$ and $c_{1},c_{2} \epsilon [0,1]$, putting $c_1 =c_2=1/2$ we get our original cost function. After doing the calculations we find that $\rho$ for this is $\frac{4(c_1+c_2\frac{F_e(1/N)}{1/n})}{2+c_1+2c_2\frac{F_e(1/N)}{1/N}}$. Now in this equation if $c_1$ (congestion coefficient) becomes 0 and for worst case we substitute $\frac{F_e(1/N)}{1/n} = 1$ then we arrive at $\rho =1$. Thus proving our claim that Braess' Paradox vanishes. Also if in this equation if $c_2$ (cost coefficient) becomes 0 then we arrive at $\rho =4/3$ the original bound by Roughgarden \cite{ref12} .\newline \qed

\section{Price Of Anarchy bound}
The price of anarchy bound for our network is  $\frac{3+\sqrt{5}}{2} \approx 2.618$. Its same as the bound given for Atomic Selfish Routing in the book \cite{ref12}. See Appendix for the proof.

\section{Conclusion, Discussions and future Work}
So we conclude with the result of reduced Braess' Paradox. And one can easily see that our model can be extended for the negligible and splittable non-atomic case as well. Our model opens several interesting future research perspectives.

\begin{enumerate}

    \item In what other game theoretical models can this \emph{buy in bulk and get discounted} phenomenon  be applied to. For examples problems like resource allocation  etc. \newline

    \item The effect of this model on undirected graphs.\newline

    \item It will be interesting to see how addition of another term say QOS(Quality of Service) affects the equations.

\end{enumerate}

\section*{Appendix: Price Of Anarchy bound}

Formally \emph{if (G,r,t)  is a network instance in which $t_e$'s congestion term is an affine function then the price of anarchy can be at most $\frac{3+\sqrt{5}}{2}$}. Let \emph{f} be the flow in the network at equilibrium and if $f^*$ is the optimal flow of the network which minimizes its \textbf{SC}{(social cost) as defined in earlier sections}. If user \emph{i} was using path $P_i$ in \emph{f} and $P_i^*$ in $f^*$ then

\begin{equation}\label{eqInequality}
\sum\limits_{e\epsilon P_i} \left( (a_{e}f_e + b_e) + u_e(r_i)  \right) \leq \sum\limits_{e\epsilon P_i^*} \left( (a_{e}(f_e+r_i) + b_e) + u_e(r_i)  \right)
\end{equation}

where $r_i$ is user \emph{i}'s flow. This equation comes from the equilibrium flow condition, definition(\ref{equilibriumCond}). Now given the same network instance and flows as for equation(\ref{eqInequality}) We have

\begin{eqnarray*}
SC(f) = \sum\limits_{i=1}^{k}r_i\left(\sum\limits_{e\epsilon P_i} a_{e}f_e+b_e + u_e(r_i)\right)  \\
\leq  \sum\limits_{i=1}^{k}r_i\left(\sum\limits_{e\epsilon P_i^*} a_{e}(f_e+r_i)+b_e + u_e(r_i)\right)\\
\leq  \sum\limits_{i=1}^{k}r_i\left(\sum\limits_{e\epsilon P_i^*} a_{e}(f_e+f_e^*)+b_e\right) + \sum\limits_{i=1}^{k}r_i\left(\sum\limits_{e\epsilon P_i^*}u_e(r_i)\right)\\
=  \sum\limits_{e\epsilon E}f_e^*\left( a_{e}(f_e+f_e^*)+b_e\right) + \sum\limits_{i=1}^{k}r_i\left(\sum\limits_{e\epsilon P_i^*}u_e(r_i)\right)\\
= SC(f^*) + \sum\limits_{i=1}^{k}r_i\left(\sum\limits_{e\epsilon P_i^*}u_e(r_i)\right)
\end{eqnarray*}

\begin{equation}\label{eqInequality2}
\Rightarrow SC(f) \leq SC(f^*) + \sum\limits_{e\epsilon E}a_{e}f_{e}f_e^*
\end{equation}

Using Cauchy-Schwartz Inequality we see that

\begin{eqnarray*}
\sum\limits_{e\epsilon E}a_{e}f_{e}f_e^* \leq \sqrt{\sum\limits_{e\epsilon E}a_{e}f_{e}^2} \sqrt{\sum\limits_{e\epsilon E}a_{e}(f_e^*)^2} \leq \sqrt{SC(f)}\sqrt{SC(f)}
\end{eqnarray*}

Combining the above result with equation(\ref{eqInequality2})we arrive at the following inequality
\begin{equation}\label{inEqPOA}
\frac{SC(f)}{SC(f^*)} - 1 \leq \sqrt{\frac{SC(f)}{SC(f^*)}}
\end{equation}

As we know $\frac{SC(f)}{SC(f^*)}$ is our \textbf{POA} defined in definition(\ref{POA}). Thus solving inequality(\ref{inEqPOA}) gives us

\begin{eqnarray*}
POA \leq \frac{3+\sqrt{5}}{2}
\end{eqnarray*}\newline\qed

\end{document}